\documentclass[%
 reprint,
 amsmath,amssymb,
 aps,
 pra,
]{revtex4-1}

\usepackage{graphicx}
\usepackage{dcolumn}
\usepackage{bm}
\usepackage{multirow}

\begin{document}

\preprint{APS/123-QED}

\title{Understanding anti-parity-time symmetric systems with a conventional heat transfer framework -- comment on ``Anti-parity-time symmetry in diffusive systems''}


\author{Lin Zhao}
	\thanks{Equal contribution to this work}
\author{Lenan Zhang}
	\thanks{Equal contribution to this work}
\author{Bikram Bhatia}
\author{Evelyn N. Wang}
	\email{enwang@mit.edu}
\affiliation{Department of Mechanical Engineering, Massachusetts Institute of Technology, Cambridge, MA 02139, USA}
\date{\today}

\begin{abstract}
Inspired by non-Hermitian physics, Li \textit{et al}. (Science 364, 170-173) theoretically predicted and experimentally demonstrated a stationary temperature profile in a diffusive heat transfer system -- seemingly indicating that heat "stops" diffusing. By analogy to the wave physics framework, the motionless and moving temperature profiles are manifestations of the anti-parity-time ($\mathcal{APT}$) symmetry and symmetry breaking states, respectively. Their experimental setup consists of two thermally coupled rings rotating in the opposite direction. At a particular rotation speed, known as the exceptional point, the $\mathcal{APT}$ symmetry of the system changes, resulting in the temperature profile switching between stationary and moving states. In fact, this seemingly unusual and exotic behavior can be elegantly captured and predicted using a conventional heat transfer framework with similarity and scaling analysis. In this work, we show the system behavior can be characterized into three zones by two widely-used dimensionless parameters on a regime map. The exceptional point, discovered using wave physics, is located precisely on the zone boundary on the regime map, indicating a balance between the contribution of thermal coupling and mechanical motion. Furthermore, the observed cessation of thermal diffusion is merely a result of the long diffusion time constant of the experimental setup. The unfamiliarity of concepts in another scientific field as well as the remarkable equivalence of the two points of view prompts this in-depth discussion of the analogy between wave physics and heat transfer. We believe that this work can help bridge the gap and promote new developments in the two distinctly different disciplines.
\end{abstract}

\maketitle



In their report \cite{Li2019}, Li \textit{et al}. studied the heat transfer in counter-moving rings that exhibits anti-parity-time ($\mathcal{APT}$) symmetry. Leveraging concepts from quantum physics and photonics, they predicted and demonstrated a critical rotation speed at the exceptional point resulting in spontaneous symmetry breaking that changes the temperature field from a motionless state to a moving state. While analogies among different scientific disciplines have led to several innovations, such as the development of photonic crystals \cite{Joannopoulos2008}, vastly different terminologies often make it challenging to appreciate the equivalence among interdisciplinary fields such as wave physics and heat transfer. Here we provide a corresponding heat transfer analysis of the same system and more explicitly show the relationship between the two seemingly distinct approaches. 

The starting point of our heat transfer analysis is the thermal energy equation \cite{Mills1999},
\begin{equation} \label{eq:1}
\rho c \frac{\partial T}{\partial t}=k \frac{\partial^2 T}{\partial x^2}-\rho c v \frac{\partial T}{\partial x}+\frac{h}{b}(T-T_{\infty})
\end{equation}

\noindent where $\rho$, $c$, and $k$ are the density, specific heat, and thermal conductivity of the medium, $v$ is the velocity of the moving medium, $h$ is the heat transfer coefficient (we follow the heat transfer convention of $h$, which has the unit $\mathrm{W/m^2/K}$), $b$ is the thickness of the medium, and $T_{\infty}$ is an arbitrary reference temperature. In the specific case of this study, the medium is the polycaprolactam (PA6) ring and $T_{\infty}$ is the temperature of the other coupled ring. Note that $h/\rho cb$, defined in this work, is equivalent to “$h$” in \cite{Li2019}.

\begin{figure*}[ht]
\includegraphics[width=6in]{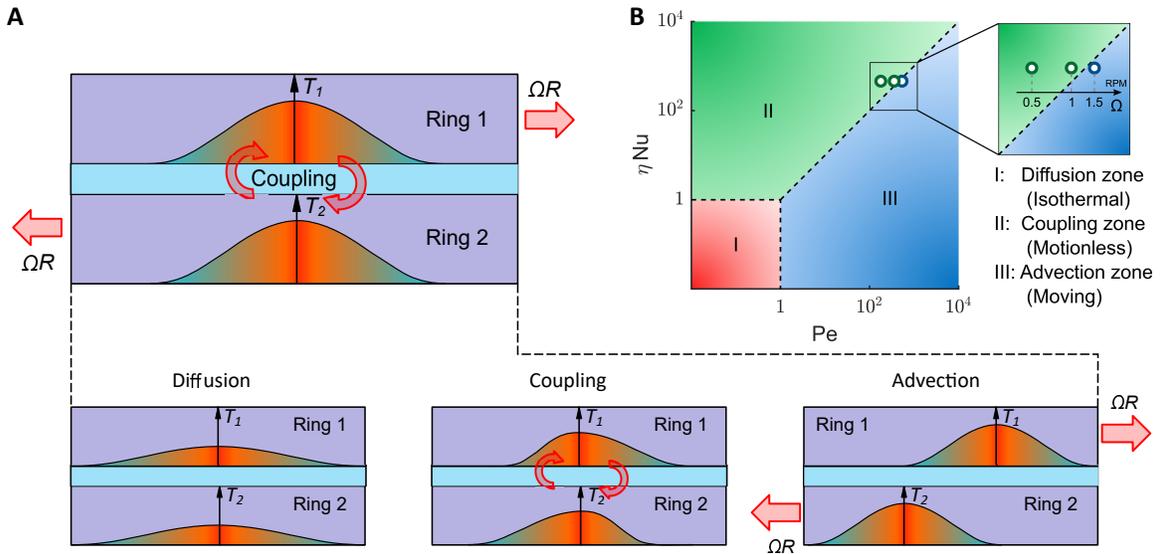}
\caption{\label{fig:1}System behavior and regime map based on two dimensionless parameters: $\mathrm{Pe}$ and $\eta\mathrm{Nu}$. (A) Schematic showing three competing heat transfer mechanisms (diffusion, coupling, and advection) occurring simultaneously in the system. Diffusion, driven by the temperature gradient in the individual ring, causes the temperature profile to smooth out and eventually become isothermal at equilibrium. Coupling, driven by the temperature difference between the two rings, induces inter-ring thermal energy exchange and reduces their temperature difference. Advection, driven by the mechanical motion of the ring, shifts the temperature profile on each ring in the direction of movement. The resultant system behavior depends on the relative importance of the three mechanisms. (B) System regime map as a function $\mathrm{Pe}$ and $\eta\mathrm{Nu}$. Corresponding experimental conditions reported by Li \textit{et al}. are shown as open circles on the map. The dashed diagonal line represents the boundary between the motionless (Zone II) and moving (Zone III) temperature fields, which is equivalent to the exceptional point showed by Li \textit{et al}. By increasing the rotation speed ($\Omega$), Li \textit{et al}. demonstrated the temperature fields changing from the motionless (Inset: green circles in Zone II) to the moving state (Inset: blue circle in Zone III). In Zone I, where $\mathrm{Pe}$ and $\eta\mathrm{Nu}$ are smaller than unity, thermal diffusion becomes dominant and the temperature field will always evolve to isothermal conditions first.}
\end{figure*}

As in a typical heat transfer analysis, Eq.~(\ref{eq:1}) is non-dimensionalized using the following length, time, and temperature scales,

\begin{equation} \label{eq:2}
x^*=\frac{x}{R},\:t^*=\frac{t}{t_{d}},\:T^*=\frac{T-T_{\infty}}{T_{\mathrm{max}}-T_{\infty}}
\end{equation}

\noindent where $R$ is the radius of the ring, $t_{d} = {R^2}/{\alpha}$ is the thermal diffusion time scale, $\alpha = {k}/{\rho c}$ is the thermal diffusivity of the ring, $T_{\mathrm{max}}$ is the maximum temperature in the system, and quantities with asterisks are dimensionless variables. Substituting Eq.~(\ref{eq:2}) into Eq.~(\ref{eq:1}), the dimensionless form of the thermal energy equation is given by,

\begin{equation} \label{eq:3}
\frac{\partial T^*}{\partial t^*}=\frac{\partial^2 T^*}{\partial t^{*2}}-\mathrm{Pe}\frac{\partial T^*}{\partial x^*}+\eta \mathrm{Nu}T^*
\end{equation}

\noindent where $\mathrm{Pe} = vR/\alpha = \Omega R^2/\alpha$ is the Peclet number, $\mathrm{Nu} = hR/k$ is the Nusselt number, and $\eta = R/b$ is the ratio of radius to thickness of the ring. These dimensionless numbers are commonly used in heat transfer analysis since they show the similarity of different systems. Specifically, the Peclet and Nusselt numbers describe the relative importance of advection (mechanical motion) and thermal coupling (between ring 1 and ring 2) over thermal diffusion, respectively, which define the dynamic similarity. $\eta$ is the aspect ratio of the ring, which defines the geometric similarity. The three competing effects in the system, \textit{i.e.}, diffusion, coupling, and advection, are depicted in Fig.~\ref{fig:1}A.

We show that the phase transition explained by the $\mathcal{APT}$ symmetry viewpoint in \cite{Li2019} can be well-predicted by dimensionless parameter analysis. According to Eq.~(\ref{eq:3}), the system characteristic is fully determined by two dimensionless parameters: $\mathrm{Pe}$ and $\eta\mathrm{Nu}$. Depending on the two dimensionless parameters, the system behavior can be divided into three zones as shown in Fig.~\ref{fig:1}B. When $\mathrm{Pe}<1$ and $\eta\mathrm{Nu}<1$ (Zone I), thermal diffusion is dominant and the temperature field will first evolve to isothermal condition regardless of thermal coupling and mechanical motion of the rings. When  $\mathrm{Pe}>1$ or $\eta\mathrm{Nu}>1$, the system behavior depends on the relative magnitude of these two parameters with a transition at $\mathrm{Pe}=\eta\mathrm{Nu}$ (the dashed diagonal line in Fig.~\ref{fig:1}B). When $\mathrm{Pe}<\eta\mathrm{Nu}$ (Zone II), thermal coupling between the two rings is dominant, which enables the “motionless” temperature field as reported in \cite{Li2019} at low rotation speeds (green circles in Fig.~\ref{fig:1}B inset). On the other hand when $\mathrm{Pe}>\eta\mathrm{Nu}$ (Zone III), advection is dominant and the temperature field moves with the ring as reported in \cite{Li2019} at high rotation speed (blue circle in Fig.~\ref{fig:1}B inset). The critical rotation speed $\Omega_{\mathrm{cr}}=1.27~\mathrm{rpm}$ corresponding to the exceptional point discovered in \cite{Li2019} is equivalent to the phase transition line between Zones II and III in Fig.~\ref{fig:1}B, represented by $\mathrm{Pe}=\eta\mathrm{Nu}$. Therefore, this regime map described by dimensionless parameters shows the same physics as the $\mathcal{APT}$ symmetry analysis \cite{Li2019}.

The system behavior predicted by the regime map (Fig.~\ref{fig:1}B) can be further validated by numerically solving Eq.~(\ref{eq:3}), where the experimental results in Figs. 4(G-I) in \cite{Li2019} were also reproduced. Boundary conditions were applied for both rings ($T^{*}_{1,2}(\theta=0)=T^{*}_{1,2}(\theta=2\pi)$), where $\theta$ is the angular location. The initial dimensionless temperature profile, $T^{*}_{1,2}(\theta,t=0)=(1+\sin{(\theta-\pi/2)})/2$, was adapted from \cite{Li2019} (Materials and Methods S3) to match the experimental conditions. The temperature peak at $\theta=\pi$ represents the hottest point on the ring, which is in contact with the high temperature copper block initially. The temperature valley at $\theta=0$ and $\theta=2\pi$ represents the coldest point on the ring, which is in contact with the low temperature copper block initially.

\begin{figure*}[ht]
\includegraphics[width=6in]{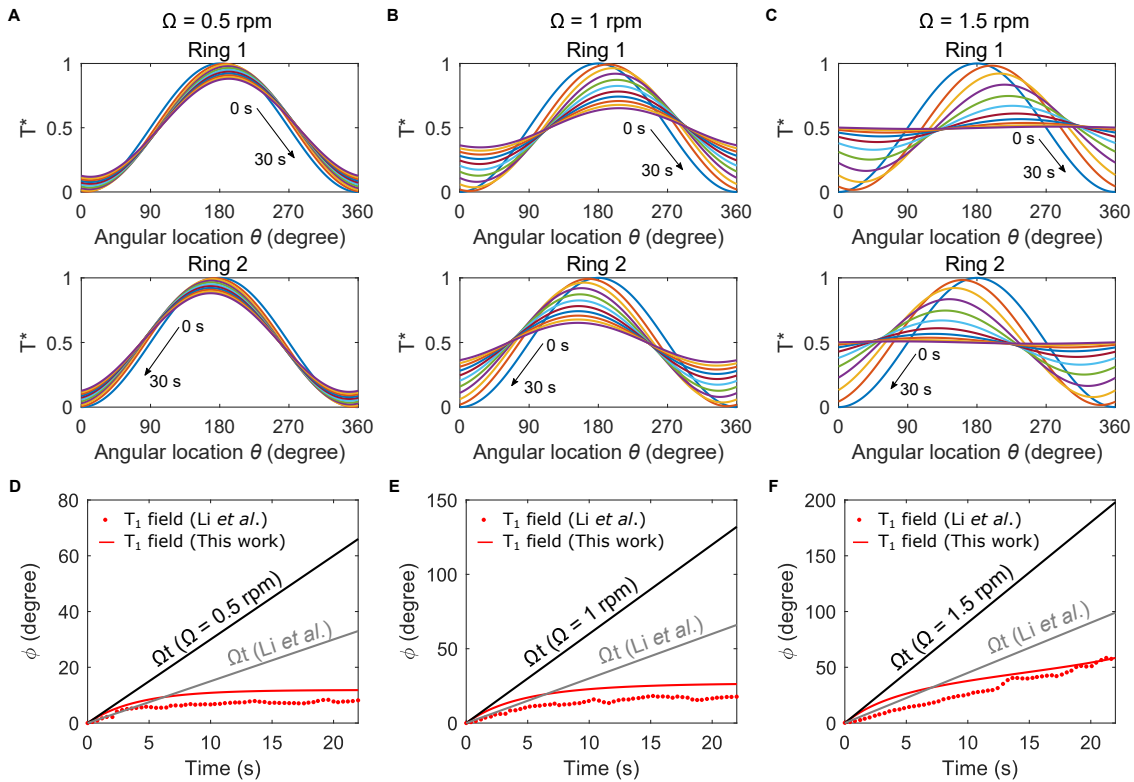}
\caption{\label{fig:2}Temperature profile evolution predicted by heat transfer analysis. (A to C) Temperature profiles of ring 1 (upper) and ring 2 (lower) at three rotation speeds, (A) 0.5 rpm, (B) 1 rpm, and (C) 1.5 rpm. Ten profiles equally spaced between $t=0$ s and $t=30$ s are shown for each ring. (D to F) Evolution of the temperature dipole angle and rotation angle of ring 1 at three rotation speeds, (D) 0.5 rpm, (E) 1 rpm, and (F) 1.5 rpm. Red dots: experimentally measured temperature dipole angles reported by Li \textit{et al}. Red curves: predicted temperature dipole angles in this work. Light gray line: incorrect rotation angles of ring 1 reported by Li \textit{et al}. Black line: correct rotation angles of ring 1 in this work.}
\end{figure*}

Figs.~\ref{fig:2}(A-C) show the temperature profile evolution of both rings (ten profiles equally spaced between 0 s and 30 s) at three rotation speeds (0.5, 1, 1.5 rpm) used in the experiments reported in \cite{Li2019}. The rotation angle $\phi$ of the temperature dipole (defined in \cite{Li2019}, Materials and Methods S4) from the initial state were then calculated from the temperature profile evolution at the corresponding rotation speeds, plotted as red curves in Figs.~\ref{fig:2}(D-F). Experimental temperature dipole angles (red dots in Figs.~\ref{fig:2}(D-F)) and rotation angles of ring 1 (light gray lines in Figs.~\ref{fig:2}(D-F)) reported in Figs. 4(G-I) in \cite{Li2019} are also included for comparison. We note an inconsistency between the rotation angles of ring 1 reported in \cite{Li2019} and in this work. We believe that this is due to an error in Figs. 4(G-I) in \cite{Li2019}, since $1~\mathrm{rpm}=6~\mathrm{deg/s}$. Overall, there is good agreement between experimentally measured temperature results reported in \cite{Li2019} and numerical results in this work, which demonstrates that heat transfer analysis using dimensionless parameters is adequate to predict the phase transition of the system.

To further demonstrate the underlying connection between the wave physics point of view proposed in \cite{Li2019} and the heat transfer analysis in this work, we summarize the equivalence between the two approaches in Table~\ref{tab:1}. Although the two points of view appear vastly different, they can both describe the same system behavior and yield identical results.

\begin{table*}[b]
\caption{\label{tab:1}%
Analogy between the wave physics point of view and conventional heat transfer
}
\begin{ruledtabular}
\begin{tabular}{lll}
& Li~\textit{et al.} & This work\\
\colrule
&&\\[-1ex]
View of problem & Anti-parity-time symmetry & Similarity and scaling analysis\\
&&\\[-1ex]
\hline
&&\\[-1ex]
\multirow{2}{*}{System characteristic} & Hamiltonian: & Dimensionless parameters:\\
& $H(k,D,h,v)$ & $f(\mathrm{Pe}, \eta\mathrm{Nu})$\\
&&\\[-1ex]
\hline
&&\\[-1ex]
\multirow{2}{*}{Motionless profile criteria} & $\mathcal{APT}$ symmetric: & Thermal-coupling dominant:\\
& pure imaginary eigenvalues & $\mathrm{Pe}<\eta\mathrm{Nu}, \eta\mathrm{Nu}>1$\\
&&\\[-1ex]
\hline
&&\\[-1ex]
\multirow{2}{*}{Moving profile criteria} & $\mathcal{APT}$ symmetry broken: & Advection dominant:\\
& non-zero real part eigenvalues & $\mathrm{Pe}>\eta\mathrm{Nu}, \mathrm{Pe}>1$\\
&&\\[-1ex]
\hline
&&\\[-1ex]
\multirow{2}{*}{Diffusing profile criteria} & - & Diffusion dominant:\\
& - & $\mathrm{Pe}<1, \eta\mathrm{Nu}<1$\\
&&\\[-1ex]
\hline
&&\\[-1ex]
\multirow{3}{*}{Phase transition location} & Exceptional point (EP): & Regime boundary:\\
& $k^2v^2_{\mathrm{EP}}=h^2$ (motion) & $\mathrm{Pe}=\eta\mathrm{Nu}$  ~~~~~~~~~(motion)\\
& ~~~~~~~-~~~~~~~ (diffusion) & $\mathrm{Pe}=1$, $\eta\mathrm{Nu}=1$ (diffusion)\\
\end{tabular}
\end{ruledtabular}
\end{table*}

\bibliography{APT_comment}

\end{document}